\documentclass[journal,10pt]{IEEEtran}
\usepackage{theorem}    \usepackage{verbatim}
 \usepackage{amssymb}  \usepackage{color}
\usepackage{bm}
\usepackage{xcolor}
\usepackage{algorithm}
\usepackage{algorithmic}

\newtheorem{T-Prob}{Transformed Problem}

\newcommand{\secref}[1]{Sec. \ref{#1}}
\newcommand{\tabincell}[2]{\begin{tabular}{@{}#1@{}}#2\end{tabular}}
\usepackage{cite}

\ifCLASSINFOpdf
  \usepackage[pdftex]{graphicx}
\else
  \usepackage[dvips]{graphicx}
\fi
\usepackage{amsmath}
\usepackage{algorithmic}

\usepackage{array}

\ifCLASSOPTIONcompsoc
  \usepackage[caption=false,font=normalsize,labelfont=sf,textfont=sf]{subfig}
\else
  \usepackage[caption=false,font=footnotesize]{subfig}
\fi

\usepackage{stfloats}
\ifCLASSOPTIONcaptionsoff
 \usepackage[nomarkers]{endfloat}
 \let\MYoriglatexcaption\caption
 \renewcommand{\caption}[2][\relax]{\MYoriglatexcaption[#2]{#2}}
\fi
\usepackage{url}

\title{A Review of Codebooks for CSI Feedback in 5G New Radio and Beyond}

\author{Ziao~Qin, Haifan~Yin
\thanks{Z. Qin and H. Yin are with Huazhong University of Science and Technology,
430074 Wuhan, China (e-mail: ziao\_qin@hust.edu.cn, yin@hust.edu.cn .}

\thanks{This work was supported by the National Natural Science Foundation of China under Grant 62071191. The corresponding author is Haifan Yin.}}

\begin{document}
\maketitle

\begin{abstract}
Codebooks have been indispensable for wireless communication standard since the first release of the Long-Term Evolution in 2009. They offer an efficient way to acquire the channel state information (CSI) for multiple antenna systems. Nowadays, a codebook is not limited to a set of pre-defined precoders, it refers to a CSI feedback framework, which is more and more sophisticated. In this paper, we review the codebooks in 5G New Radio (NR) standards. The codebook timeline and the evolution trend are shown. Each codebook is elaborated with its motivation, the corresponding feedback mechanism, and the format of the precoding matrix indicator. Some insights are given to help grasp the underlying reasons and intuitions of these codebooks. Finally, we point out some unresolved challenges of the codebooks for future evolution of the standards. In general, this paper provides a comprehensive review of the codebooks in 5G NR and aims to help researchers understand the CSI feedback schemes from a standard and industrial perspective.
\end{abstract}

\begin{IEEEkeywords}
MIMO; FDD; codebook; CSI; 5G NR.
\end{IEEEkeywords}
\section{introduction}

Since the first release of NR technical specifications, R15, in late 2017, the fifth generation (5G) mobile communication is being deployed all over the world. To meet the ever-growing user requirements, the 5G NR specification keeps evolving, and R17 is finalized in 2022. According to the 3rd Generation Partner Project (3GPP), R18 will be officially referred to as ``5G Advanced''. In fact, 5G NR technology evolves from Long-Term Evolution (LTE), which jointly provides the overall 5G radio-access solution with NR \cite{dahlman5GNRbook}. 

Multiple-Input Multiple-Output (MIMO) has been an integral technology to improve system performance since 4G LTE R8 released in 2009. In 5G NR, this technology has evolved to massive MIMO \cite{2014ErikMMIMO} with an increasing scale of the antenna array.  Massive MIMO provides higher transmission diversity, higher spatial multiplexing gain, and higher transmission directivity. Hence, higher spectral efficiency and more reliability can be achieved \cite{2010MazattaMIMO}. Particularly, the key to high transmission directivity brought by massive MIMO is beamforming, which enables multi-user spatial multiplexing. To achieve accurate beamforming, Channel State information (CSI) is the indispensable premise. At the base station (BS) side, the downlink (DL) CSI can be acquired by the feedback information from the users (UEs), i.e., CSI report \cite{jose2011pilot}. Note that CSI report is more indispensable in frequency division duplex (FDD) mode than time division duplex (TDD) mode \cite{2016Emil}. The reported CSI enables the BS to calculate the precoding matrix for beamforming and user scheduling. {In 3GPP standards, the CSI report process is achieved by the configuration of the codebook and the feedback of the codewords.} At first, a codebook refers to a set of pre-defined precoders, a.k.a., codewords, and the UEs feed back the indices of the codewords to the base station. With the development of the standard nowadays, the meaning of codebook extends to the whole CSI report mechanism, which helps the base station compute the precoding matrix with the feedback from the UEs.  

The CSI report framework includes the procedure of a particular CSI reference signal (CSI-RS) transmitted by the BS and a series of feedback information from the UEs. Even though 5G NR evolves from LTE, the CSI acquisition framework in NR is quite different. Particularly in LTE, the CSI acquisition framework is coupled with the transmission modes (TMs). For example, a codebook-based feedback mode is defined in TM6, also known as the close-loop scheme. At the same time, the open-loop scheme is also supported in LTE, which means no CSI report is needed for precoding. In 5G NR, however, CSI report framework is decoupled with the TM and relies on the CSI report configurations instead. In this way, better flexibility and scalability for CSI report are achieved. 

More specifically, the CSI report framework configuration consists of two parts, i.e., report resources setting and report type setting \cite{3gpp214r17}. The report resources setting specifies the periodic report manner and the occupied bandwidth part (BWP) according to different usages of the reference signal. For example, the CSI reference signal specializes in CSI calculation \cite{3gpp213}. And the report type is configured based on report resources configuration. It mainly reports the CSI indicators and the corresponding codebook configuration. Particularly, the layer indicator (LI) and the rank indicator (RI) specify the optimum layer with best quality and the maximum number of transmission layers, respectively. Correspondingly, the precoding matrix indicator (PMI) is utilized for the base station to reconstruct or calculate the DL precoders. 

The essence of CSI report framework lies in the codebook design which determines the obtained precoding matrix from CSI feedback. The corresponding PMI indicates the specific channel characteristics with a chosen codebook scheme. In fact, since 4G, the codebooks have been evolving towards characterizing more detailed channel information with less time-frequency overhead. The number of supported types of codebook has increased over time to six in 5G NR R17 to accommodate different system requirements and to maintain backward compatibility. {Recently, novel methods, such as machine learning \cite{2022machinefeedback}, joint spatial division and multiplexing (JSDM) \cite{2019JSDM} and computer vision \cite{2022cvfeedback}, are utilized to improve the accuracy of the CSI feedback.}

{From an industrial point of view, some researchers reviewed the CSI acquisition mechanism and technology. The authors in \cite{liu2012downlink} reviewed the downlink CSI transmission and acquisition in R10 LTE-Advanced. In \cite{wang2018spatial}, the authors reviewed the channel estimation method based on the spatial-wideband effect in a millimeter-wave MIMO system. In \cite{ramireddy2022enhancements}, the authors discussed the multi-beam operation method and high-resolution codebook for 5G NR multi-user MIMO (MU-MIMO). The enhancement of Type II Codebook in the UE mobility scenario for 5G NR was discussed in \cite{onggosanusi2022enhancing}. A review of uplink and downlink transmission operations from R15 to R17 and the emerging technology of massive MIMO in R18 is made by the authors in \cite{2023R18review}. These research works offer some perspectives on high-accuracy CSI feedback. However, there is a lack of review about the PMI reporting, evolution trend, and performance comparison of the codebooks.}

In this paper, we focus on discussing the codebook evolution, the corresponding PMI report, and the precoding matrix mapping in 5G NR. We first elaborate the codebook evolution timeline and the future developing trend. The relationship and comparison between codebooks are presented. We also explain the physical meanings of important parameters to describe a codebook. Then a thorough analysis of the PMI format, the mapping relationship from the PMI to the precoding matrix, and the PMI report strategies of the codebooks are given. {Besides, a performance comparison among the codebooks is reviewed, including the number of report beams, the subband report manner, and the feedback overhead. The evaluations of the codebook performance vs. feedback overhead are also given.} In the end, we discuss some open problems in codebooks for 5G advanced and the sixth Generation (6G) wireless technology, including the support of high mobility, cell-free massive MIMO, and ultramassive MIMO. To the best of our knowledge, this is the first paper that provides an overview of the practical codebooks widely adopted by industry. Since academia and industry diverge a lot nowadays, this paper serves as a bridge over the increasing gap between academia and industry in the perspective of CSI feedback, with the hope of bring the practical limitations and the ideas of industry to the attention of academic researchers.  

The following of the paper is organized in eight sections. First, we review the codebook evolution history and elaborate the physical meaning of the configured parameters in \secref{sec1}. Then, from \secref{sec2} to \secref{sec5}, each codebook is reviewed with details of the PMI format and how the next generation NodeB (gNB) calculates the precoding matrix from the reported PMI. In \secref{sec6}, the codebooks for 5G beyond are discussed. In the end, \secref{sec7} concludes the paper.

\section{Codebook evolution}\label{sec1}
\begin{figure}[!ht]
\centering
\includegraphics[width=0.45\textwidth]{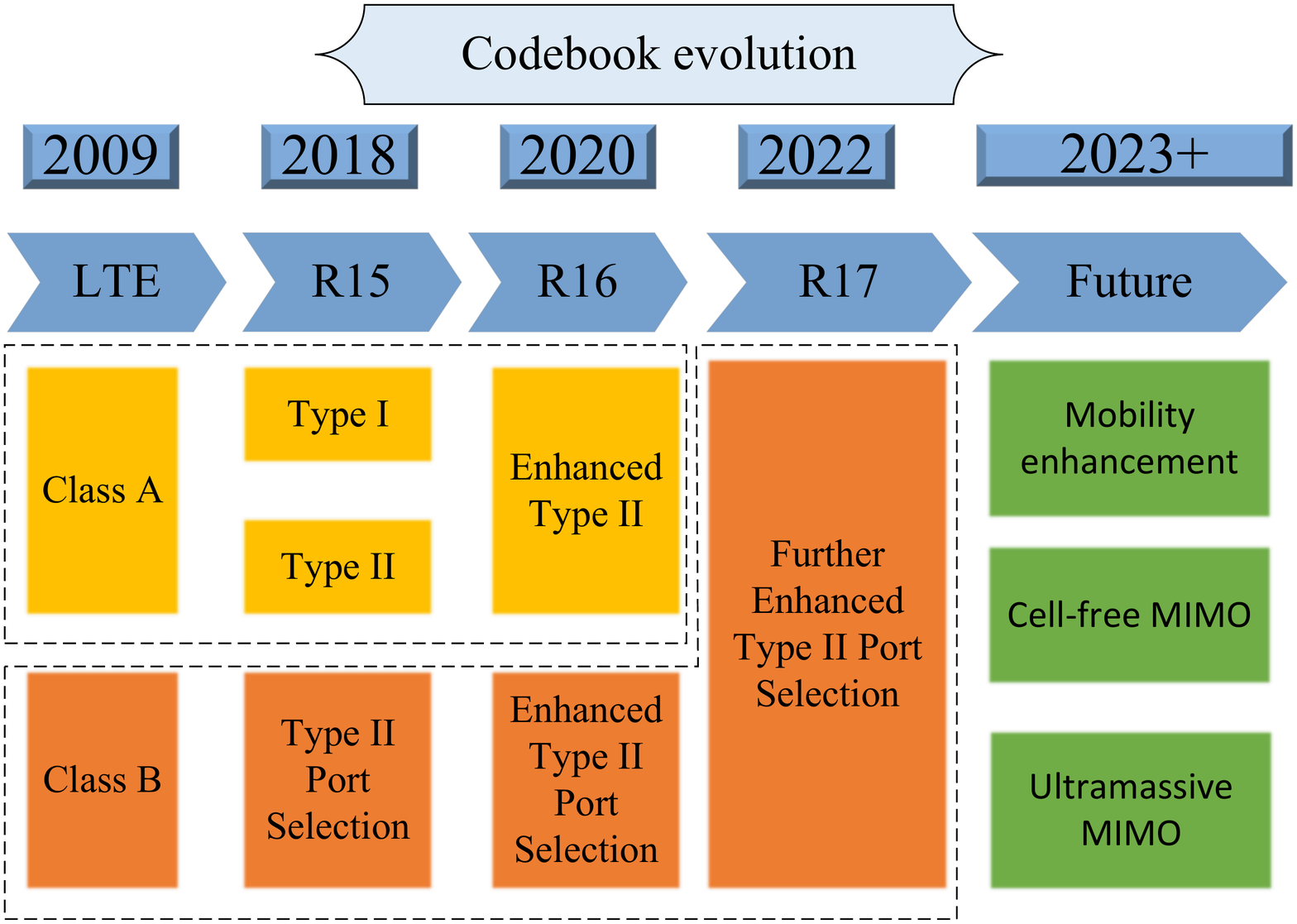}
\caption{Codebook evolution from LTE to 5G NR and beyond.}
\label{fig_Codebook evolution}
\end{figure}
Since the first release of LTE in 2009, codebooks have been evolving due to the advances of multi-antenna technology and the growing performance requirements. An illustration of the codebook evolution is shown in Fig. \ref{fig_Codebook evolution}. In the beginning, two different codebooks are supported. Class A Codebook is based on a classical closed-loop feedback. The precoder is based on the feedback from the UEs and comes from a discrete Fourier transform (DFT) matrix \cite{brady2013beamspace}. And the pilot sequences are not precoded at the BS. On the contrary, Class B Codebook relies on precoded pilots and the UE selects the precoder after estimating the precoded effective channel, e.g., selecting the precoder index (port number) corresponding to the largest amplitude of the pilot-based channel estimation. The BS may choose the reported precoder for DL data transmission. Note that the precoders here may not be limited to DFT vectors like in Class A Codebook. 

In the first version of 5G NR released in 2018, new types of codebook were introduced as a derivation of Class A and Class B. Type I Codebook and Type II Codebook evolve from Class A Codebook. Type II Codebook aims to provide more details of the spatial signature than Type I Codebook at the cost of heavier feedback overhead. {Meanwhile, Type II Port Selection Codebook inherits the basic idea of Class B Codebook and supports a multi-beam report.} In 2020, 5G NR R16 introduced Enhanced Type II Codebook and Enhanced Type II Port Selection Codebook. The most significant characteristic of Enhanced Type II Codebook is the support of subband-wise calculation of the PMI while the feedback overhead is balanced through a joint spatial and frequency domain compression. Such a compression is enabled by a larger number of BS antennas and a broader supported bandwidth. And Enhanced Type II Port Selection Codebook is evolved from Type II Port Selection Codebook likewise. In the recent R17, Further Enhanced Type II Port Selection Codebook is proposed in order to further improve the performance of Enhanced Type II Port Selection Codebook with the help of partial reciprocity between the uplink (UL) and downlink (DL) channel \cite{yin2021codebook,3gpp:36.897}. In 5G Advanced and 6G, we believe some other enhanced features will be enabled by future codebooks, for example, to support high mobility transmission, cell-free MIMO, ultramassive MIMO, etc.

In the following of the paper, we provide a comprehensive review of each codebook from the perspective of the PMI report mechanism and the calculation of the corresponding precoding matrix.
Before elaborating on the details of each codebook, some terminologies and symbols to describe the codebook are explained below: 
\begin{itemize}
\item layers: The streams in MIMO-enabled spatial multiplexing, i.e., transmitting signals simultaneously in the same time/frequency resources.
\item subband: Several consecutive resource blocks (RBs). The bandwidth of a subband may be configured as 4, 8, 16 RBs, etc. 
\item beam: {It means a certain spatial direction, normally corresponding to a column vector from a one-dimensional or two-dimensional DFT matrix, when the antenna array topology is an uniform linear array (ULA) or an uniform planar array (UPA), respectively.}
\item antenna port: This terminology is not related to a physical ``port" anymore. The symbols transmitted on the same antenna port can be assumed to share the same effective channel. 
\item $\upsilon$: The layer limitation configured by the gNB and indicated by RI.
\item $N_{\rm{AP}}$: The number of antenna ports at the gNB.
\item $N_1,N_2,O_1,O_2$: $N_1$ and $N_2$ denote the number of antennas elements in the horizontal and vertical direction, respectively. $O_1$ and $O_2$ are the oversampling factors in the horizontal and vertical direction, respectively.
\item $N_g$: The number of antenna panels.
\item $L$: The number of reported beams in a certain codebook.
\item $N_3$: The number of subbands in a BWP.
\end{itemize}

{The major characteristic of each codebook lies in how it maps the precoding matrix from the reported PMI. Overall, the precoding matrix consists of beam information and phase information, which rely on the reported beam indicators and phase indicators. To be more specific, the UE utilizes the received signal to search for the optimal codeword based on the precoder form constraint. Then it reports the codeword index as the beam indicator and quantizes the corresponding phase as the phase indicator. In the following sections, we focus on introducing the PMI format and the calculation of the precoding matrix.}

\section{Type I Codebook}\label{sec2}
In R17, two sub-types of Type I Codebook are supported, i.e., Type I Codebook with Single-Panel and Type I Codebook with Multi-Panel. The main difference between the two codebooks is the number of the supported transmit antenna panels. First, we discuss Type I Codebook with Single-Panel.  
\subsection{Type I Codebook with Single-Panel}
\begin{figure}[!ht]
\centering
\includegraphics[width=0.45\textwidth]{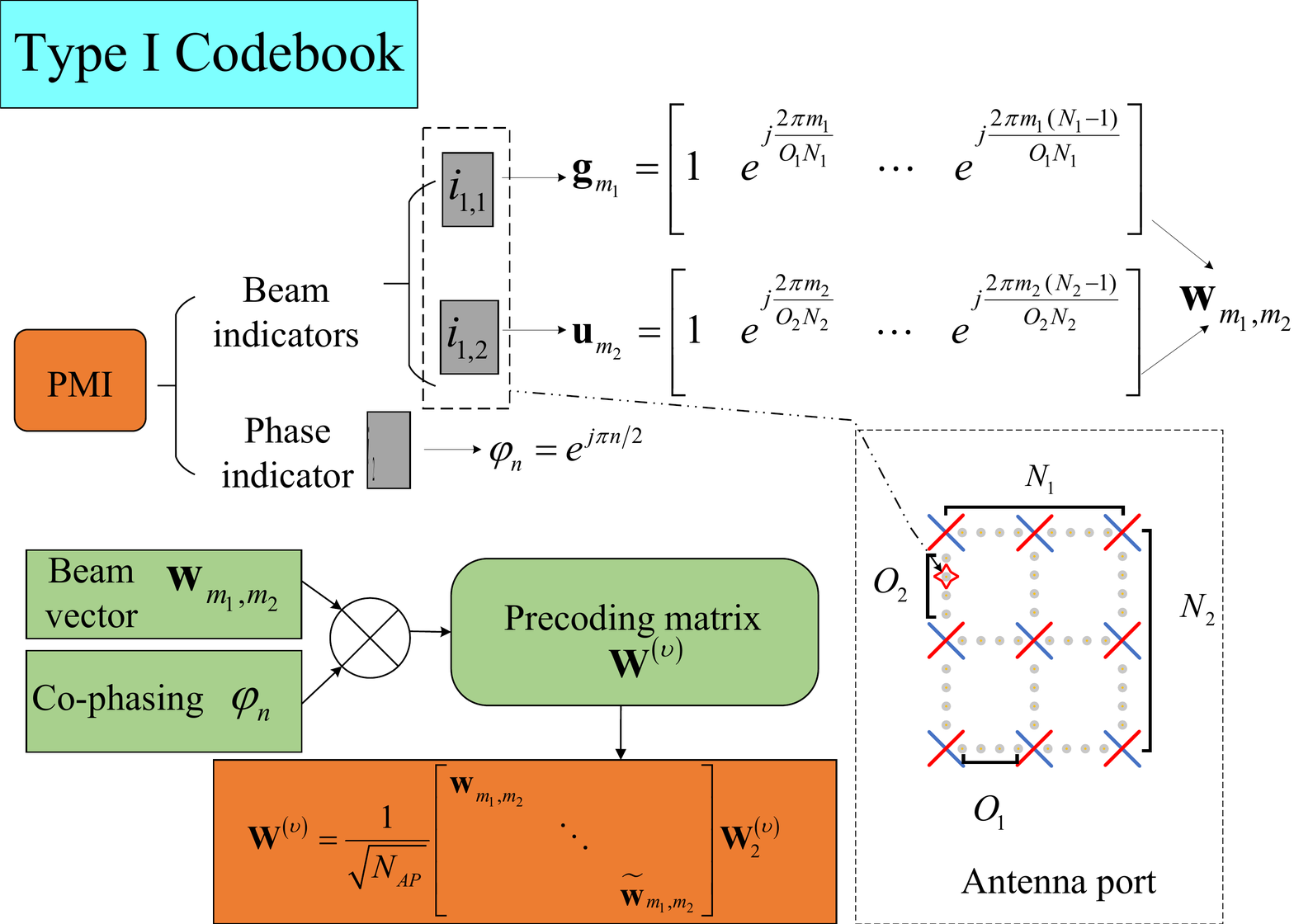}
\caption{The 2D antenna port structure, PMI format and precoding matrix calculation of Type I Codebook.}
\label{fig_Type I Codebook}
\end{figure}

Type I Codebook with Single-Panel is relatively straightforward as the reported PMI reflects the information of a single beam, including the beam selection and the co-phasing information among the dual-polarized antennas. Under the assumption of an uniform planar array (UPA) at the gNB as in \cite{3gpp901}, the chosen beam is selected from the set of 2D DFT vectors with spatial oversampling, indicated by $\left( {{N_1},{N_2},{O_1},{O_2}} \right)$. These parameters are specified by Table 5.2.2.2.1-2 in \cite{3gpp214r17}. Fig. \ref{fig_Type I Codebook} demonstrates the physical meanings of the PMI. Define the PMI vector as ${\mathbf{I}}{\text{ = }}\left[ {\begin{array}{*{20}{c}}{{{\mathbf{I}}_1}}&{{{\mathbf{I}}_2}} \end{array}} \right]$, where ${{\bf{I}}_1}$ reports the chosen beam information and ${{\bf{I}}_2}$ indicates the corresponding phase information, respectively. ${{\bf{I}}_1}$ includes two indicators $i_{1,1}$ and $i_{1,2}$. The indicator $i_{1,1}$ maps the horizontal beam index $m_1$ and the indicator $i_{1,2}$ maps the vertical beam index $m_2$. The second part of the PMI ${{\bf{I}}_2}=i_2$ maps the co-phasing information by ${\varphi _n} = {e^{j{{\pi n} \mathord{\left/{\vphantom {{\pi n} 2}} \right.\kern-\nulldelimiterspace} 2}}}$. We should note that $n$ is binary, except when $\upsilon = 1$, $n\in \left\{ {0,1,2,3} \right\}$. When the layer limitation $\upsilon\le2$, the beam choice is indicated together by $i_{1,1},i_{1,2},i_{2}$, otherwise by $i_{1,1},i_{1,2}$.

The 2D antenna array structure and the PMI format of Type I Codebook are illustrated in Fig. \ref{fig_Type I Codebook}. The 2D beam ${\bf{w}}_{m_1,m_2}$ is a Kronecker product of the vertical beam ${{\bf{u}}_{m_2}}$ and the horizontal beam ${{\bf{g}}_{m_1}}$. And the neighboring 2D beam ${{{\widetilde {\bf{w}}}_{m_1,m_2}}}$ consists of a different horizontal beam ${\widetilde {\bf{g}}_{m_1}}$ and the same vertical beam with ${\bf{w}}_{m_1,m_2}$. The indices $m_1$ and $m_2$ are indicated by $i_{1,1}$ and $i_{1,2}$, respectively. Fig. \ref{fig_Type I Codebook} also demonstrates how to calculate the precoding matrix from the reported PMI and the equation is given  
\begin{equation}
    {{\bf{W}}^{\left( \upsilon  \right)}} = \frac{1}{{\sqrt {{N_{{\rm{AP}}}}} }}\left[ {\begin{array}{*{20}{c}}
{{{\bf{w}}_{{m_1},{m_2}}}}&{}&{}\\
{}& \ddots &{}\\
{}&{}&{{{\widetilde {\bf{w}}}_{{m_1},{m_2}}}}
\end{array}} \right]{\bf{W}}_2^{\left( \upsilon  \right)}
\nonumber
\end{equation}
{The precoding matrix ${\bf{W}}^{\left( \upsilon  \right)}$ is made of the beam vector and phasing information, which are indicated by the 2D beam ${\bf{w}}_{m_1,m_2}$ and the subband phase $\phi_n$, respectively.}
This procedure is relatively straightforward and only includes one beam and the co-phasing information. More specifically, the 2D beam ${\bf{w}}_{m_1,m_2}$ is frequency-irrelevant and reported in wideband mode. In contrast, the co-phasing matrix ${\bf{W}}_2^{\left( \upsilon  \right)}$ is frequency-dependent and needs to be reported per subband. The column vector of ${\bf{W}}_2^{\left( \upsilon  \right)}$ characterizes the co-phasing information of each layer, which is specified in Table 5.2.2.2.1-(5-12) in \cite{3gpp214r17}. {However, when $N_{\rm{AP}}>16$ and $\upsilon \in \left\{3,4\right\}$, the matrix ${\bf{W}}_2^{\left( \upsilon  \right)}$ also indicates a beam choice between the beam ${\bf{w}}_{m_1,m_2}$ and another beam ${{{\widetilde {\bf{w}}}_{m_1,m_2}}}$ which is defined in Table 5.2.2.2.1 of \cite{3gpp214r17}.}

Evolved from LTE, this codebook is rather simple and works well in strong line of sight (LoS) scenarios. {The number of reported coefficients is smaller compared to other codebooks.} The computational complexity of calculating the precoding matrix from the PMI is the lowest. Due to the concise structure of Type I Codebook with Single-Panel, the layer limitation can be eight. However, the performance of this codebook is limited, particularly so in multipath scenarios, since only one beam is used for signal transmission.

\subsection{Type I Codebook with Multi-Panel}
This codebook may be applied when the antenna array at the gNB consists of multiple antenna panels instead of one. In order to facilitate the implementation, the number of antenna ports $N_{\rm{AP}}$ is limited to the set $\left\{ {8,16,32} \right\}$ and the layer limitation descends to $\upsilon \le 4$. In this codebook, additional co-phasing information needs to be reported.

Compared to Type I Codebook with Single-Panel, a new indicator vector ${\bf{i}}_{1,4}$ is introduced in Type I Codebook with Multi-Panel. The dimension of the vector ${\bf{i}}_{1,4}$ is associated with the number of antenna panels $N_g$ and the codebook mode $C_m$, varying from one to three. It indicates the inter-panel co-phasing information and the dual-polarization co-phasing information. Other indices in ${\bf{I}}_1$ are consistent with Type I Codebook with Single-Panel. However, the indicator ${{\bf{I}}_2}$ is reported in a different manner. When $C_m=2$, it consists of three indices, ${i_{2,0}}$,  ${i_{2,1}}$ and ${i_{2,1}}$. Otherwise, it only includes one index $i_2$, as in Type I Codebook with Single-Panel. 

The precoding matrix ${{\bf{W}}^{\left( \upsilon  \right)}}$ of Type I Codebook with Multi-Panel is similar to Type I Codebook with Single-Panel. The main difference lies in the co-phasing matrix ${\bf{W}}_2^{\left( \upsilon  \right)}$, which is jointly indicated by ${{\bf{I}}_2}$ and ${\bf{i}}_{1,4}$. It relies on the parameter combination $\left( {{N_g},{C_m},\upsilon } \right)$, which is specified in Table 5.2.2.2.2-1 of \cite{3gpp214r17}. Particularly, the additional co-phasing information is quantified by ${a_p} = {e^{j{\pi  \mathord{\left/
 {\vphantom {\pi  4}} \right.
 \kern-\nulldelimiterspace} 4}}}{e^{j{{\pi p} \mathord{\left/
 {\vphantom {{\pi p} 2}} \right.
 \kern-\nulldelimiterspace} 2}}}$ and ${b_n} = {e^{ - j{\pi  \mathord{\left/
 {\vphantom {\pi  4}} \right.
 \kern-\nulldelimiterspace} 4}}}{e^{j{{\pi n} \mathord{\left/
 {\vphantom {{\pi n} 2}} \right.
 \kern-\nulldelimiterspace} 2}}}$. The indices $n\in{n_0,n_1,n_2}$ are indicated by ${\bf{I}}_2$ and $p\in{p_1,p_2}$ are indicated by ${{\bf{i}}_{1,4}}$. 

In general, the two sub-types of Type I Codebook mentioned above are both able to provide the beam information and the co-phasing information. It is particularly applicable in a single-user MIMO (SU-MIMO) scenario. Besides, Type I Codebook lays the foundation of the other codebooks in subsequent releases of 5G NR. However, the drawbacks of Type I Codebook are also explicit. Due to the large bandwidth in 5G NR, the channels of subbands differ a lot. Type I Codebook only allows for one spatial beam, which will be used in the whole BWP. Hence, it has limited capability to characterize the channel with multipath. As a result, the spectral efficiency performance of Type I Codebook is unsatisfactory, especially in massive MIMO. Therefore, other types of codebooks are naturally proposed as enhancements, such as Type II Codebook.

\section{Type II Codebook}\label{sec3}
Type II Codebook is first proposed in 5G NR R15 as an upgrade of Type I Codebook, in order to better characterize the multi-path channel. One of the most significant improvement of Type II Codebook is the support of multiple beams. Each beam and its corresponding coefficient reflect a path with a certain angle. And up to four beams can be reported in Type II Codebook. As a result, this codebook outperforms Type I Codebook in most scenarios, nevertheless, at the cost of increased feedback overhead.
\subsection{PMI format}
\begin{figure}[!ht]
\centering
\includegraphics[width=0.45\textwidth]{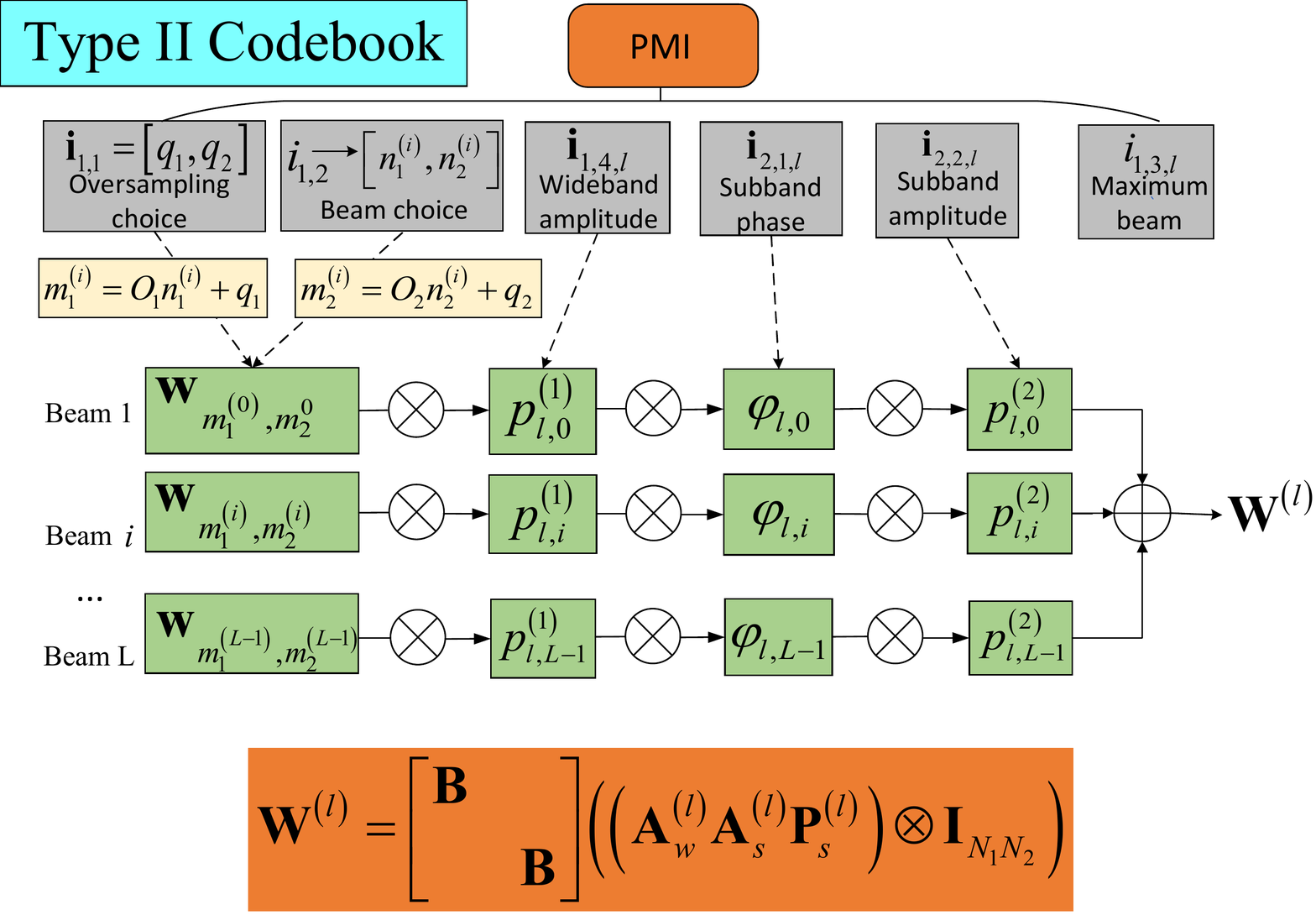}
\caption{The PMI format and the precoding matrix of Type II Codebook at layer $l$.}
\label{fig_Type II Codebook}
\end{figure}

The PMI report for Type II Codebook turns out to be more complicated than Type I Codebook. As a tradeoff between the performance and the feedback overhead / complexity, the layer limitation $\upsilon$ is 2. Fig. \ref{fig_Type II Codebook} demonstrates the PMI format, which covers four kinds of beam information, i.e., the beam choice, the beam with the maximum amplitude, the beam amplitudes and the beam phases. The chosen $L\in \left\{2,3,4\right\}$ beams are indicated by ${{\bf{i}}_{1,1}}$ and ${i_{1,2}}$. We should note that all layers share the same beam choice. The indicator ${\bf{i}}_{1,1}$ contains two indices $q_1,q_2$, where ${q_1}$ and ${q_2}$ map the oversampling parameter in the horizontal and vertical direction, respectively. The indicator ${i_{1,2}}$ indicates how to choose $L$ beams from the DFT vector set of size $N_1N_2$. The value of ${i_{1,2}}$ varies from $0$ to $C_{{N_1}{N_2}}^L-1$, where $C_{{N_1}{N_2}}^L$ represents the number of possibilities of selecting different $L$ beams from all ${N_1}{N_2}$ beams. 
In general, the gNB is equipped with dual-polarized antennas. In Type II Codebook, the same set of $L$ beams are shared for both polarizations. As a result, $2L$ beam coefficients corresponding to $L$ chosen beams are reported. These coefficients include the amplitudes and the phases. In order to reduce the complexity, only the phase information are reported in a subband manner, while the amplitudes can be reported in subband manner or wideband manner ( the reported amplitude for a certain beam is identical for all the subbands in the whole BWP), depending on configuration.

The wideband amplitude indicator ${{\bf{i}}_{1,4,l}} $ is a vector with $2L$ entries, which are denoted by $k_{l,i}^{{\rm{(1)}}}$, where $i \in \left\{ {0, \cdots 2L - 1} \right\}$ indicates the beam index and $k_{l,i}^{{\rm{(1)}}} \in \left\{ {0,1, \ldots ,7}\right\}$. The wideband amplitude of beam $i$ at layer $l$ is computed by $p_{l,i}^{({\rm{1}})} = {\left( {{1 \mathord{\left/
 {\vphantom {1 {\sqrt 2 }}} \right.
 \kern-\nulldelimiterspace} {\sqrt 2 }}} \right)^{7 - k_{l,i}^{({\rm{1}})}}}$. 
The phase information reported in every subband are indicated by ${\bf{i}}_{2,1,l}$. Its element $c_{l,i}$ quantizes phases in a N-phase shift keying (N-PSK) manner as ${e^{j2\pi {{{c_{l,i}}} \mathord{\left/
 {\vphantom {{{c_{l,i}}} {{N_{{\rm{psk}}}}}}} \right.
 \kern-\nulldelimiterspace} {{N_{{\rm{psk}}}}}}}}$, where $N_{\rm{psk}} \in \left\{4,8\right\}$. The indicator ${i_{1,3,l}}$ maps the index of the beam with the maximum amplitude at layer $l$.

In fact, subband amplitude report can be supported in Type II Codebook and it is indicated by a binary parameter $I_{s}$. Specifically, $I_s=1$ means that subband amplitude report is enabled, while $I_s=0$ means it does not. If $I_s=1$, an additional indicator vector ${{\bf{i}}_{2,2,l}}$ is reported to quantize the subband amplitude information with its entries being $k_{l,i}^{\left( 2 \right)} \in \left\{0,1\right\}$. The reported subband amplitude is thus $p_{l,i}^{({\rm{2}})} = {\left( {{1 \mathord{\left/
 {\vphantom {1 {\sqrt 2 }}} \right.
 \kern-\nulldelimiterspace} {\sqrt 2 }}} \right)^{1 - k_{l,i}^{({\rm{2}})}}}$. 

\subsection{PMI report compression}
Type II Codebook supports multiple beams and subband-wise report. As a result, the feedback overhead becomes heavier compared to Type I Codebook. To balance the report accuracy and the overhead, Type II Codebook introduces a PMI report compression mechanism. The coefficients of beam with the strongest amplitude $k_{l,i_l^*}^{\left( 1 \right)}$ and the corresponding phase ${c_{l,i_l^*}}$ will not be reported, where the beam index $i_l^*$ is indicated by ${i_{1,3,l}}$. When Type II Codebook is configured to wideband mode, only the non-zero wideband amplitude $k_{l,i}^{{\rm{(1)}}}$ and the corresponding phase $c_{l,i}$ are reported to the gNB. The number of non-zero coefficients of each layer is $M_{\rm{nz}}^l < 2L$. If subband mode is configured, the subband coefficients report is slightly different. The ${M_{\rm{vr}}^l}$ stronger subband coefficients are phase-quantized in a N-PSK manner, where $N_{\rm{psk}} \in {4,8}$. And the remaining ${M_{{\rm{nz}}}^l} - {M_{{\rm{vr}}}^l}$ non-zero subband coefficients are phase-quantized with $N_{\rm{psk}} = 4$. The rest $2L - {M^l_{{\rm{vr}}}}$ subband coefficients are not reported, since they are very close to zero. In general, the core idea of PMI report compression lies in feeding back the information of the predominant beams, and the feedback overhead is reduced by ignoring the weak beams.

\subsection{Precoding matrix calculation}
The precoding matrix calculation in Type II Codebook is different from Type I Codebook. The main difference is that Type II Codebook supports multiple beams. Fig. \ref{fig_Type II Codebook} shows how to map the precoding matrix from the PMI in Type II Codebook.  

In general, the precoding matrix ${\bf{W}}^{\left(l\right)}$ is a weighted summation of multiple beams and is calculated by 
\begin{equation}
    {{\bf{W}}^{\left( l \right)}} = \left[ {\begin{array}{*{20}{c}}
{\bf{B}}&{}\\
{}&{\bf{B}}
\end{array}} \right]\left( {\left( {{\bf{A}}_w^{\left( l \right)}{\bf{A}}_s^{\left( l \right)}{\bf{P}}_s^{\left( l \right)}} \right) \otimes {{\bf{I}}_{{N_1}{N_2}}}} \right)
\nonumber
\end{equation}
For each beam, four types of beam information are reported, i.e., the beam choice ${\bf{w}}_{m_1^{(i)},m_2^{(i)}}$, the wideband amplitude $p_{l,i}^{({\rm{1}})}$, the subband phase $c_{l,i}$ and the subband amplitude $p_{l,i}^{{\rm{(2)}}}$. In Fig. \ref{fig_Type II Codebook}, the chosen $L$ beams are denoted by ${\bf{B}}$. The block matrix $\text{diag}\left\{{\bf{B}}, {\bf{B}} \right\}$ is introduced to represent the beams for both polarizations. The matrix ${\bf{B}}$ is composed of $L$ beams and each beam ${\bf{w}}_{m_1^{(i)},m_2^{(i)}}$ is similar to the vector ${{\bf{w}}_{m_1,m_2}}$ in Type I Codebook. However, the indices $m_1^{\left( i \right)},m_2^{\left( i \right)}$ are mapped from  ${\bf{i}}_{1,1}$ and $i_{1,2}$ as illustrated in Fig. \ref{fig_Type II Codebook}. And the beam choice indices $n_1^{\left( i \right)}$,  $n_2^{\left( i \right)}$ are calculated from $i_{1,2}$ through the algorithm in Sec. 5.2.2.2.3 of \cite{3gpp214r17}. The wideband amplitude of each layer is defined by a diagonal matrix ${\bf{A}}_w^{\left( l \right)}$ and its element $p_{l,i}^{({\rm{1}})}$ is indicated by ${\bf{i}}_{1,4,l}$. The wideband amplitude matrix ${\bf{A}}_w^{\left( l \right)}$ is always reported. 
The diagonal matrix ${\bf{A}}_s^{\left( l \right)}$ is the subband amplitude matrix, which is valid only if subband mode is supported. The diagonal elements of ${\bf{A}}_s^{\left( l \right)}$ are indicated by ${\bf{i}}_{2,2,l}$. Correspondingly, the subband phase information is characterized by ${\bf{P}}_s^{\left( l \right)}$. The elements of ${\bf{P}}_s^{\left( l \right)}$ are mapped from ${\bf{i}}_{2,1,l}$. The matrix $I_{N_1N_2}$ is an $N_1N_2 \times N_1 N_2$ identity matrix.

Overall, Type II Codebook shows many significant improvements, especially the support of subband amplitude and multiple beams. As a result, the CSI feedback is more accurate, which facilitates the gNB to cancel inter-user interference and allocate resources. 
This is also why Type II Codebook is more suitable for multi-user MIMO (MU-MIMO) than Type I Codebook. Even though Type II Codebook introduces a PMI report compression scheme to reduce the overhead, the feedback still scales with the bandwidth and the number of UEs. This problem is particularly acute in FDD mode with large number of gNB antennas. Nowadays, the increasing number of antennas and wider bandwidth call for new codebooks with high accuracy and low feedback overhead. Fortunately, a  better-performing codebook called ``Enhanced Type II Codebook" is proposed in 5G NR R16. 

\section{Enhanced Type II Codebook}\label{sec4}
The codebooks discussed before were proposed in 5G NR R15. With the evolution of 5G NR, frequency-sensitive and multi-path channel environment requires a codebook with better performance by capturing both the spatial domain and the frequency domain structures of the channel. Hence, Enhanced Type II Codebook is proposed in 5G NR R16 as an Upgrade of Type II Codebook. It is particularly suitable in a multipath scattering environment with diverse angle spread and delay spread, while the UE is capable of complex signal processing. 

\begin{figure}[!ht]
\centering
\includegraphics[width=0.45\textwidth]{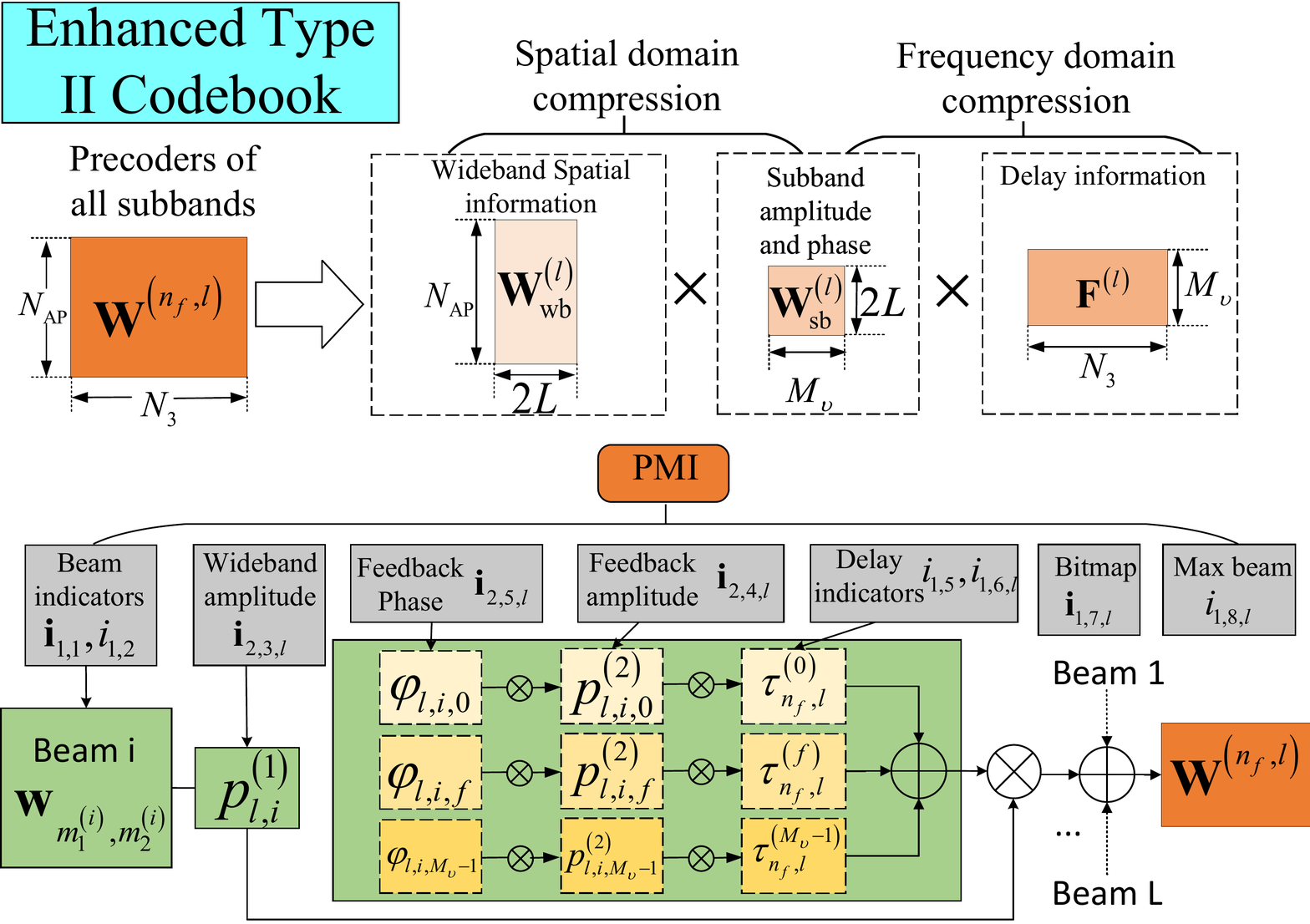}
\caption{The compression in spatial and frequency domain and the PMI format of Enhanced Type II Codebook.}
\label{fig_Enhanced Type II Codebook}
\end{figure}

The most significant merit of Enhanced Type II Codebook lies in feedback reduction in spatial and frequency domain. This is enabled by the channel sparsity in both spatial and frequency domains in wideband massive MIMO. Fig. \ref{fig_Enhanced Type II Codebook} gives a demonstration of the feedback overhead compression. In the spatial domain, $L$ beams are chosen to characterize the angular structure of the channel like in Type II Codebook. However, the subband amplitude is always reported in Enhanced Type II Codebook. In the frequency domain, a delay matrix ${\bf{F}}^{\left(l\right)}$ is introduced to map the phase information of all $N_3$ subbands with $M_\upsilon\le N_3$ basis vectors. Hence, the subband amplitude and phase of all beams of all $N_3$ subbands are reported in ${\bf{W}}^{\left(l\right)}_{\rm{sb}}$ with the help of $M_\upsilon$ IDFT vectors. Due to the DFT-based compression in spatial domain and the IDFT-based compression in frequency domain, Enhanced Type II Codebook has a reduced feedback overhead compared with its predecessor. 

According to Table 5.2.2.2.5-1 in \cite{3gpp214r17}, eight compression configurations, denoted by the parameter combination $\left( {L,{p_\upsilon},\beta } \right)$, for Enhanced Type II Codebook  are supported. 
The number of  basis vectors in frequency domain is calculated by ${M_\upsilon } = \left\lceil {{p_\upsilon }\frac{{{N_3}}}{R}} \right\rceil$, where $p_v \in \left\{ 1/4, 1/8\right\}$ is the number of average basis vectors used per subband in frequency domain. $\beta \in\left\{1/4, 1/2, 3/4\right\}$ is the feedback overhead compression ratio from the full dimension to the reduced dimension. The parameter $R$ is either one or two, depending on the higher-layer configurations. Therefore, in spatial domain and frequency domain, a total of $LM_\upsilon$ basis vectors are utilized to characterize the precoding matrix.

\subsection{PMI format}
The PMI format in Enhanced Type II Codebook is more complicated than Type II Codebook. As illustrated in Figure \ref{fig_Enhanced Type II Codebook}, the PMI format includes the beam indicators  ${\bf{i}}_{1,1},i_{1,2}$, the delay indicators $i_{1,5},i_{1,6,l}$, the bitmap indicator $i_{1,7,l}$, the strongest beam indicator $i_{1,8,l}$, the wideband amplitude indicator ${{\bf{i}}_{2,3,l}}$, the feedback amplitude indicator ${{\bf{i}}_{2,4,l}}$ and the feedback phase indicator ${{\bf{i}}_{2,5,l}}$.

On one hand, the beam indicators are similar to the ones in Type II Codebook. The beam selection is mapped by ${\bf{i}}_{1,1},i_{1,2}$ like Type II Codebook. The wideband amplitude indicator ${{\bf{i}}_{2,3,l}}$ consists of two coefficients, ${k_{l,0}^{\left( 1 \right)}}$ and ${k_{l,1}^{\left( 1 \right)}}$. They quantize the wideband amplitude in each polarization direction with 4 bits according to the mapping relationship in Table 5.2.2.2.5-2 of \cite{3gpp214r17}. The quantified wideband amplitude at each polarization direction is denoted by $p_{l,0}^{\left( 1 \right)}$ and $p_{l,1}^{\left( 1 \right)}$. Compared with the wideband amplitude indicator ${\bf{i}}_{1,4,l}$ in Type II Codebook, the amplitude quantization in Enhanced Type II Codebook increases from 3 bits to 4 bits. Moreover, the subband beam information is always available in Enhanced Type II Codebook. It is reported in angle-delay domain. The coefficients $k_{l,i,f}^{\left( 2 \right)}$ of ${{\bf{i}}_{2,4,l}}$ quantize the feedback amplitude $p_{l,i,f}^{\left( 2 \right)}$ with 3 bits, outperforming the 1-bit quantization of the subband amplitude in Type II Codebook. Corresponding to the feedback amplitude $p_{l,i,f}^{\left( 2 \right)}$, the coefficients $\phi_{l,i,f}$ of indicator ${{\bf{i}}_{2,5,l}}$ quantize the feedback phase ${c_{l,i,f}}$ in a 4PSK manner. The indicator ${i_{1,8,l}}$ records the index of the strongest subband coefficient at layer $l$, similar to the indicator $i_{1,3,l}$ in Type II Codebook. 

On the other hand, due to the compression in frequency domain and the report of delay information, several new indicators $i_{1,5},i_{1,6,l},{\bf{i}}_{1,7,l}$ are introduced. The subband amplitude and phase information is reported in $M_\upsilon$ dimension instead of $N_3$, due to the frequency domain compression. The frequency basis vectors are determined by a vector ${{\bf{n}}_{3,l}} \in {\mathbb{C}^{1 \times {M_\upsilon }}}$. Each element of this vector, denoted by $n_{3,l}^{\left( f \right)} \in \left\{ {0,1 \cdots {N_3} - 1} \right\},f \in \left\{ {0, \cdots {M_\upsilon } - 1} \right\}$, indicates the delay information of the corresponding frequency basis vector through the relationship $\tau _{{n_f},l}^{\left( f \right)} = {e^{j{{2\pi {n_f}n_{3,l}^{\left( f \right)}} \mathord{\left/
 {\vphantom {{2\pi {n_f}n_{3,l}^{\left( f \right)}} {{N_3}}}} \right.
 \kern-\nulldelimiterspace} {{N_3}}}}}$ is the subband index. The vector ${{\bf{n}}_{3,l}}$ is computed based on the indicators $i_{1,5},i_{1,6,l}$ that are fed back by the UE, according to the algorithm in Sec. 5.2.2.2.5 of \cite{3gpp214r17}. Denote the index of the strongest frequency basis vector at the layer $l$ by $f^*_l$. The frequency basis vector ${\bf{n}}_{3,l}$ is reorganized with respect to $f^*_l$ such that $n_{3,l}^{\left( f \right)} = \left( {n_{3,l}^{\left( f \right)} - n_{3,l}^{\left( {f_l^*} \right)}} \right)\bmod {N_3}$. Thus, $n_{3,l}^{\left( {f_l^*} \right)} = 0$ after remapping. Likewise, the frequency basis vector index $f$ is reorganized with respect to $f^*_l$ such that ${f} = \left( {{f} - f_l^*} \right)\bmod {M_\upsilon }$, and therefore, $f^*_l=0$. 

\subsection{PMI report compression}
Although the problem of feedback overhead is alleviated by IDFT based frequency domain compression in Enhanced Type II Codebook, the PMI report still consumes valuable time-frequency resources. In order to further reduce the overhead, some PMI compression mechanisms are introduced. 

First, the indices of the strongest beam at the layer $l$ are denoted by $i^*_l$. 
The coefficients of ${{\bf{i}}_{2,4,l}},{{\bf{i}}_{2,5,l}}$ corresponding to $i^*_l,f^*_l$, as well as the wideband amplitude ${{\bf{i}}_{2,3,l}}$  with indices equal to $\left\lfloor {{{i_l^*} \mathord{\left/
 {\vphantom {{i_l^*} L}} \right.
 \kern-\nulldelimiterspace} L}} \right\rfloor$ 
are not reported. Then, similar to Type II Codebook, only non-zero coefficients of ${\bf{i}}_{2,4,l}$ and ${\bf{i}}_{2,5,l}$ are reported. The indicator ${\bf{i}}_{1,7,l}$ serves as a bitmap with size $1 \times 2L M_\upsilon$ in order to show whether the UE reports the corresponding coefficients in ${\bf{W}}_{\text{sb}}^\text{(l)}$ or not. Since some values in ${\bf{W}}_{\text{sb}}^\text{(l)}$ are negligible, this bitmap will help reduce the feedback overhead. 
The number of reported coefficients of all layers is denoted by ${M_{{\rm{nz}}}} = \sum\limits_{l = 1}^\upsilon  {M_{{\rm{nz}}}^l}$. The number of non-zeros coefficients ${M_{\rm{nz}}^l}$ is equal to the summation of the coefficients of the bitmap indicator ${\bf{i}}_{1,7,l}$ at layer $l$. 
As a result, $2L\upsilon {M_v} - {M_{\rm{nz}}}$ coefficients of ${{\bf{i}}_{2,4,l}},{{\bf{i}}_{2,5,l}}$ are not reported, where $\upsilon$ is the number of layers. Since in each layer, only relative values with respect to the coefficient with the maximum amplitude are needed for feedback, the number of all reported coefficients $k_{l,i,f}^{\left( 2 \right)},{c_{l,i,f}}$ is thus ${M_{\rm{nz}}} - \upsilon$. 

\subsection{Precoding matrix calculation}
In general, the precoding matrix calculation in Enhanced Type II Codebook has a lot in common with Type II Codebook. The precoding matrix ${{\bf{W}}^{\left( {{n_f},l} \right)}}$ is similar to ${{\bf{W}}^{\left( l \right)}}$ in Fig. \ref{fig_Type II Codebook} The main difference lies in the frequency domain compression and the mapping of the delay information. 
Fig. \ref{fig_Enhanced Type II Codebook} demonstrates the relationship between the PMI and the precoding matrix ${{\bf{W}}^{\left( {{n_f},l} \right)}}$. The beam selecting matrix $\bf{B}$ is consistent with Type II Codebook. However, the wideband amplitude matrix ${\bf{A}}_w^{\left( l \right)}$ is different, as it is composed of a block diagonal matrix with the two blocks reflecting the wideband amplitudes for both polarization instead of reusing the same set of wideband amplitudes among the two polarizations as in Type II Codebook. The reconstruction of the subband phase and amplitude is also quite different from Type II Codebook, because of the frequency domain compression with IDFT basis vectors. The amplitude and phase information of all subbands are transformed to angle-delay domain, quantized, and fed back to the gNB. Then the gNB reconstructs the information by reverse transformation with the quantized coefficients. 

Generally speaking, Enhanced Type II Codebook is more sophisticated than Type II Codebook. Despite the complexity, Enhanced Type II Codebook shows great potential in improving system spectral efficiency. The detailed PMI report in Enhanced Type II Codebook characterizes much more channel structure information, especially in the delay domain. The key lies in the exploitation of the multipath angle-delay structure of wideband massive MIMO by means of DFT and IDFT transformations. Thanks to the feedback overhead reduction in frequency domain, the maximum number of layers in Enhanced Type II Codebook increases to four. And the maximum number of beams $L$ increases from four to six compared to Type II Codebook. In fact, higher frequency band and larger antenna arrays are given great expectations for 5G NR and beyond. In such case, the angle and delay structure of the channel is more obvious and should be captured by the codebooks in order to facilitate the CSI feedback. No doubt that Enhanced Type II Codebook is a good choice in this circumstance. However, the feedback overhead of Enhanced Type II Codebook is still a serious problem, especially when the number of antennas and the bandwidth are large. Achieving more accurate CSI feedback with less overhead is an everlasting effort of industry. 

\begin{figure}[!ht]
\centering
\includegraphics[width=0.45\textwidth]{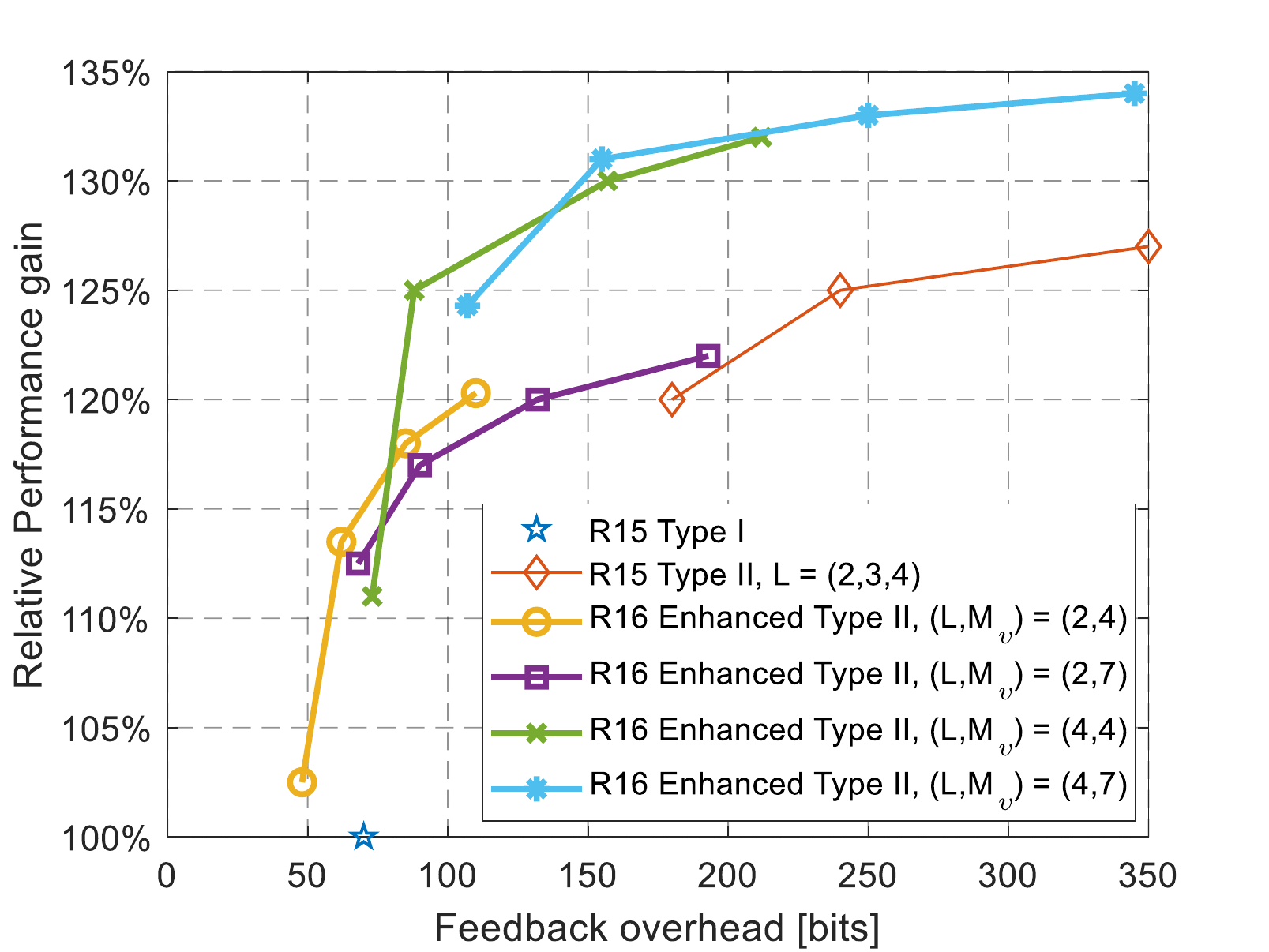}
\caption{{The performance of Enhanced Type II Codebook vs. feedback overhead, MU-MIMO, 32 ports, rank = 1, resource utilization (RU) $\approx 70\%$, $\beta  = \left\{ {\frac{1}{8},\frac{1}{4},\frac{1}{2},\frac{3}{4}} \right\}$.}}
\label{fig_codebook_performance_non_port}
\end{figure}
{According to the technical report \cite{Huawei96}, the relative gains of throughput for Enhanced Type II Codebook is demonstrated in Fig. \ref{fig_codebook_performance_non_port}. And four different feedback overhead compress ratios $\beta  = \left\{ {\frac{1}{8},\frac{1}{4},\frac{1}{2},\frac{3}{4}} \right\}$ are evaluated. The parameter $\left(L,M_\upsilon\right)$ stands for the number of the spatial-frequency basis vectors.  Learned from Fig. \ref{fig_codebook_performance_non_port}, Enhanced Type II Codebook achieves better performance gains over the former codebooks as well as lighter feedback overhead. And the number of spatial-frequency basis vectors $\left(L,M_\upsilon\right)$ play an important role in feedback overhead and performance. Note that more evaluation results under different parameter configurations can be found in \cite{Erisson97,ZTE97}.} 
\section{Port selection codebooks}\label{sec5}
A category of codebook called port selection codebooks is also supported in 3GPP standards, starting with Type II Port Selection Codebook introduced in R15. For ease of exposition, the codebooks discussed before are referred to as non-port selection codebooks in our paper. The main difference between the port selection codebooks and the previously described codebooks lies in the beam selection mechanisms. More specifically, in the non-port selection codebooks, the UE finds the spatial beams by computing the inner product between the DL CSI or precoders and the 2D DFT vectors with oversampling.  One or several strong beams are then reported by the UE. In port selection codebooks however, the gNB transmits precoded reference signal (pilot) with different precoders, where each precoder represents a certain beam and is associated with an antenna port. The UE selects several antenna ports by pilot-based measurements and reports the corresponding coefficients. As a result, the beams are determined by antenna port selection. In port selection codebooks, all $N_{\rm{AP}}$ antenna ports are grouped by a port sampling parameter $d$. Then, the beam selection is indicated by a binary port choice. 

\begin{figure}[!ht]
\centering
\includegraphics[width=0.45\textwidth]{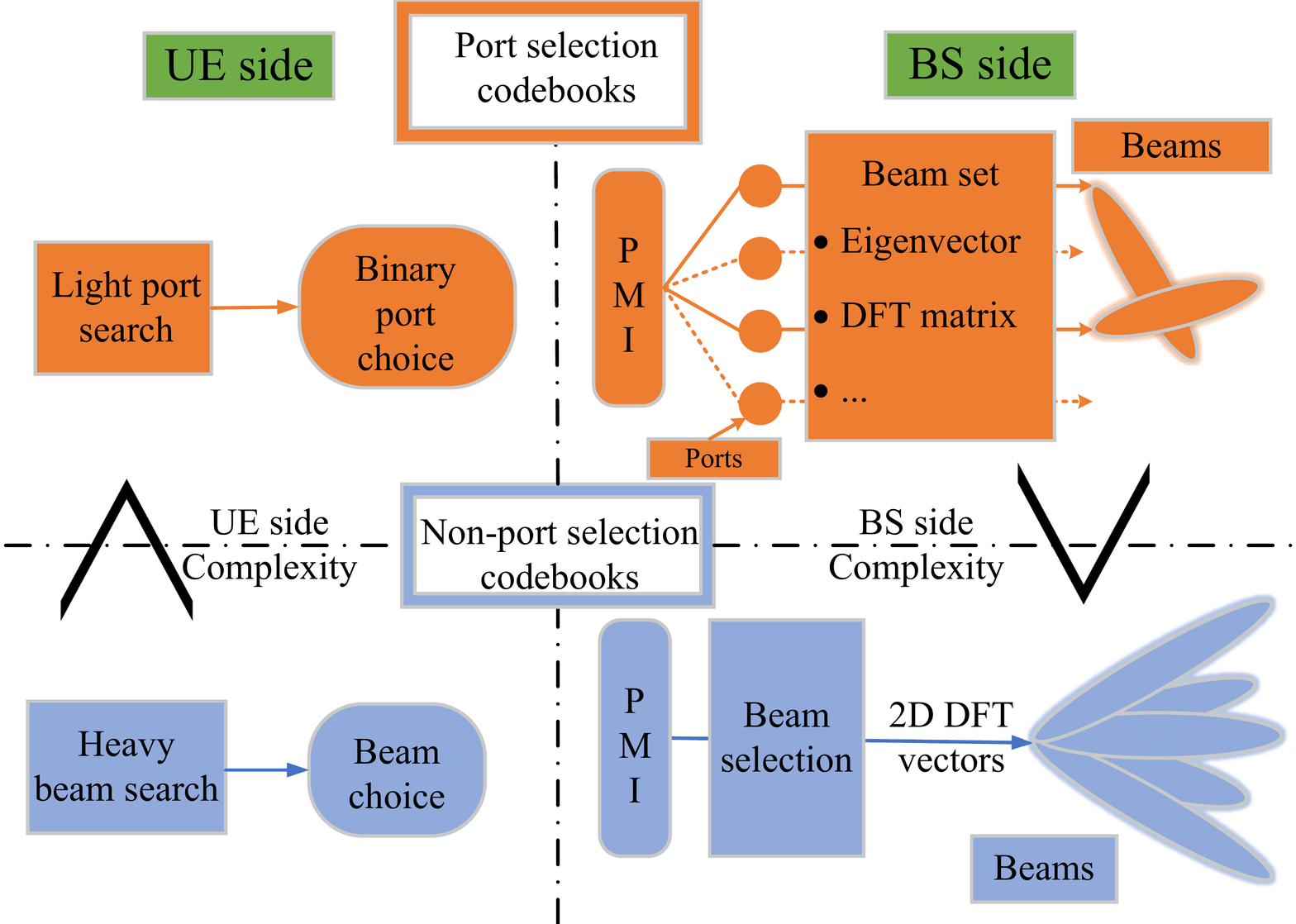}
\caption{CSI report of port selection codebooks and non-port selections codebook at the BS side and the UE side.}
\label{fig_Port selection}
\end{figure}

Fig. \ref{fig_Port selection} compares the differences between port selection codebooks and non-port selection codebooks. 

In general, the core idea of port selection codebooks lies in the fact that the UE reports a port selection decision other than a beam, and the UE is not aware of the specific beam related to a certain antenna port. After the gNB receives the reported port choice, it finds the beams corresponding to the selected antenna ports, and then reconstructs the DL precoder or CSI with the port-related quantized coefficients reported by the UE. Note that the beams are not limited to the 2D DFT vectors. It may also take the form of the eigenvectors of the channel covariance matrix, which generally outperforms the DFT vectors. Such kind of beams is enabled by the low-rankness property of the channel covariance matrix \cite{2013YinJSAC, 2013JSDM}, which facilitates the compression of the CSI using channel statistics.  

The advantages of port selection codebooks are twofold. First, the form of beams is decoupled with the UE feedback. Hence, the topology of the antenna array at gNB is no longer limited to UPA and the beams are more flexible to accommodate different antenna typologies and algorithms. On the contrary, in non-port selection codebooks, the gNB and the UE assume the beams to be 2D DFT vectors only, which may not work well under other antenna typologies than UPA. 
Second, the computation complexity is reduced at the UE side in exchange for extra beam calculation complexity at the gNB side. This is due to the binary port selection decision rather than the complex 2D beam searching procedure in non-port selection codebooks.

In 5G NR R17, three port selection codebooks are supported, i.e., Type II Port Selection Codebook, Enhanced Type II Port Selection Codebook and Further Enhanced Type II Port Selection Codebook, which will be discussed below. 

\subsection{Type II and Enhanced Type II Port Selection Codebook}
These two codebooks are proposed together with the corresponding non-port selection codebooks in R15 and R16, respectively. We focus on analyzing the port selection indicators of both codebooks.

First, the port sampling parameter $d$ is configured by the gNB. The indicator ${i_{1,1}}$ denotes the selected port sample group at each polarization. This indicator is different from the one in non-port selection codebooks. The value of ${i_{1,1}}$ varies from zero to $\left\lceil {{{{N_{\rm{AP}}}} \mathord{\left/
 {\vphantom {{{N_{\rm{AP}}}} {2d}}} \right.
 \kern-\nulldelimiterspace} {2d}}} \right\rceil  - 1$. The port selection index $q^{\left(i\right)}$ is mapped by $i_{1,1}$ as $q^{\left(i\right)} = {i_{1,1}}d + i$. 
 Finally, the $q^{\left(i\right)}$-th entry of the reported port selection vector ${\bf{w}}_{q^{\left(i\right)}} \in {\mathbb{C}^{\frac{{N_{\rm{AP}}}}{2} \times 1}}$ is one and the rest  are zero.

The remaining indicators and the precoding matrix calculation of these two codebooks are consistent with the corresponding non-port selection codebooks. Therefore, the PMI of these two codebooks can be obtained in a way similar to the corresponding non-port selection codebooks, and the details are omitted.

\subsection{Further Enhanced Type II Port Selection Codebook}
This port selection codebook is first supported in the recent 5G NR R17. Its most intriguing characteristic lies in the exploitation of the partial angle-delay reciprocity of the channel in FDD massive MIMO. Even though the complete channel reciprocity does not hold in FDD, the frequency-irrelevant parameters, e.g., the multipath angle and delay distributions of the downlink and uplink channels are very close. Such a property is exploited in Further Enhanced Type II Port Selection Codebook and the feedback overhead is reduced. The core idea of Further Enhanced Type II Port Selection Codebook is elaborated in \cite{yin2021codebook}. This codebook is enabled by a joint spatial-frequency domain precoding scheme for the transmission of the downlink CSI-RS. {The joint spatial-frequency precoders are also referred to as ``wideband precoders''. They are computed based on the uplink channel estimates, and the partial reciprocity is exploited therein. The choices of the wideband precoders can be flexible depending on different ways of implementation \cite{yin2021codebook}. The specific form of wideband precoders includes, however not limited to, the DFT vectors and the eigenvectors of the joint spatial and frequency domain channel covariance matrix.} 

Thanks to the partial channel reciprocity, Further Enhanced Type II Port Selection Codebook has the potential to achieve better system performance with less feedback coefficients, and the computational complexity at the UE side is also greatly reduced. 

Further Enhanced Type II Codebook extends the number of beams to ${{\alpha {N_{{\rm{AP}}}}} \mathord{\left/
{\vphantom {{\alpha {N_{{\rm{AP}}}}} 2}} \right.
\kern-\nulldelimiterspace} 2}$, where $\alpha$ is the ratio of the chosen antenna ports to the total antenna ports. The maximum reported beams is 6 according to Table 5.2.2.2.7-1 in \cite{3gpp214r17}. However, in frequency domain, the number of frequency basis vectors $M_\upsilon\in \left\{1,2\right\}$ is smaller than  in Enhanced Type II Port Selection Codebook.

The PMI format of Further Enhanced Type II Port Selection Codebook is similar to  Enhanced Type II Port Selection Codebook. However, it may report more beams compared to its predecessor. In frequency domain, the PMI report of Further Enhanced Type II Port Selection Codebook is quite different. Particularly, the frequency basis indicating vector  ${{\bf{n}}_3} \in \mathbb{C}^{1\times M}$ is defined like in Enhanced Type Port Selection Codebook, however it is identical across layers, rather than layer-dependent. A new indicator $i_{1,6}$ reflects the non-zero values of ${{\bf{n}}_3}$. 

\begin{figure}[!ht]
\centering
\includegraphics[width=0.45\textwidth]{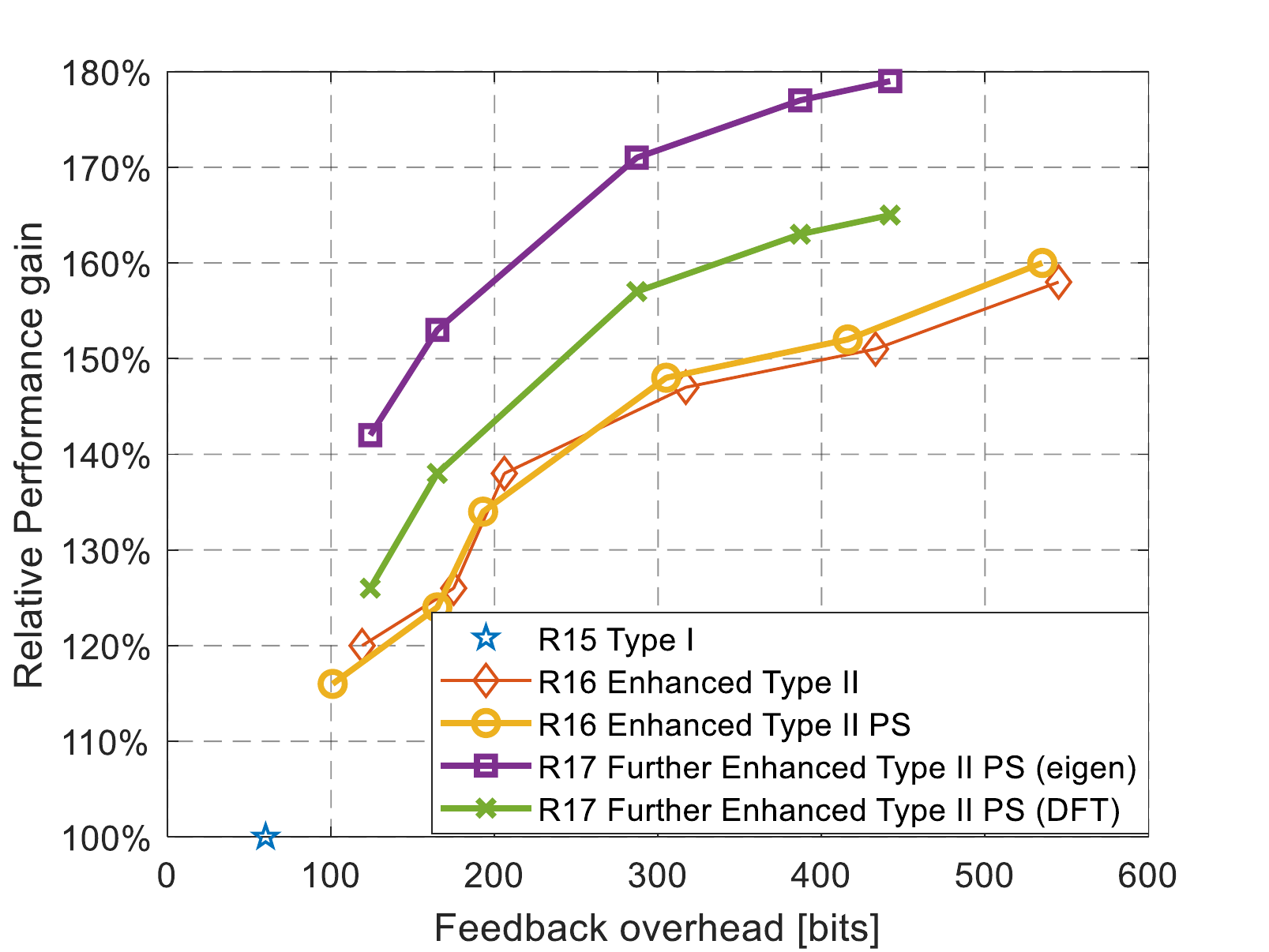}
\caption{{The performance of Port Selection (PS) Codebook vs. feedback overhead, MU-MIMO, 32 ports, rank = 2, RU $\approx 70\%$.}}
\label{fig_codebook_performance_port}
\end{figure}
{According to the technical report \cite{Huawei104e}, Further Enhanced Type II Port Selection Codebook is evaluated in terms of the relative throughput gains compared to the former codebooks in Fig. \ref{fig_codebook_performance_port}. Five parameter configurations of codebooks \cite{3gpp214r17} are evaluated, except Type I Codebook. The numerical results demonstrate that Further Enhanced Type II Port Selection Codebook outperforms the other codebooks due to the exploitation of the partial reciprocity in the spatial-frequency domain. Moreover, the eigen-based Further Enhanced Type II Port Selection Codebook achieves a higher gain than the DFT-based one. For more details and simulation results of Further Enhanced Type II Codebook, one may refer to the technical reports \cite{Erisson106e,ZTE105e}.}

Generally speaking, port selection codebooks leads to a more flexible beam set than non-port selection codebooks. To be more specific, the beams can be chosen from DFT vectors, or the eigenvectors of covariance matrix, etc. It is not limited to a certain antenna array topology. On the contrary, the non-port selection codebooks generally assume a UPA or ULA topology at the gNB, and the beams are generated from DFT vectors. However, the port selection codebooks require a more intelligent gNB algorithm to find the proper beams based on limited information, e.g., the partial reciprocity. In the non-port selection codebooks, since the UEs have the DL channel estimation, the beams are readily obtained with DFT transformations. 

\section{Codebooks for the future}\label{sec6}
\begin{table*}[!ht]
\centering \protect\protect\caption{Comparison of Codebooks in 5G NR R17}
\label{tab codebook performance}
\begin{tabular}{|c|c|c|c|c|}
\hline
 Codebook type & \tabincell{c}{Number of beams} & \tabincell{c}{Subband quantization manner} & Feedback overhead & Complexity\tabularnewline
\hline
Type I (Single-Panel) & 1 & phase & $\left\{ {2 + {N_3},3 + {N_3}} \right\}$ & ${\cal O}\left( {2 N_1N_2  \upsilon } \right)$
\tabularnewline
\hline
Type I (Multi-Panel) & 1 &  phase & $\left\{ {6 + {N_3},7 + {N_3},8 + {N_3}} \right\}$ & ${\cal O}\left( {2N_gN_1N_2\upsilon } \right)$
\tabularnewline
\hline
Type II & $\left\{2,3,4\right\}$ & \tabincell{c}{phase and \\amplitude (1bit)} & $\begin{array}{*{20}{l}}
{2 + \upsilon  + \left( {{N_3} + 1} \right)\sum\limits_{l = 1}^\upsilon  {M_{{\rm{nz}}}^l} ,{I_s} = 0;}\\
{2 + \upsilon  + \sum\limits_{l = 1}^\upsilon  {\left( {2{N_3}M_{{\rm{vr}}}^l + M_{{\rm{nz}}}^l} \right)} ,{I_s} = 1}
\end{array}$ & $ {{\cal O}}\left( {2\upsilon {L}{N_1}{N_2}} \right)$
\tabularnewline
\hline
Type II Port Selection & $\left\{2,3,4\right\}$ &\tabincell{c}{phase and \\amplitude (1bit)} & $\begin{array}{*{20}{l}}
{1 + \upsilon  + \left( {{N_3} + 1} \right)\sum\limits_{l = 1}^\upsilon  {M_{{\rm{nz}}}^l} ,{I_s} = 0;}\\
{1 + \upsilon  + \sum\limits_{l = 1}^\upsilon  {\left( {2{N_3}M_{{\rm{vr}}}^l + M_{{\rm{nz}}}^l} \right)} ,{I_s} = 1}
\end{array}$ & ${{\cal O}}\left( {2\upsilon {L}{d}} \right)$
\tabularnewline
\hline
Enhanced Type II & $\left\{2,4,6\right\}$ & \tabincell{c}{phase, delay and \\amplitude (3bit)\\with compression} & $\begin{array}{l}
2 + \upsilon  + 2LM{}_\upsilon \upsilon  + 2{M_{\rm{nz}}},{N_3} \le 19;\\
3 + \upsilon  + 2LM{}_\upsilon \upsilon  + 2{M_{\rm{nz}}},{N_3} > 19
\end{array}$ & ${\cal O}\left( {2\upsilon L{ M_\upsilon }{N_1}{N_2}} \right)$
\tabularnewline
\hline
\tabincell{c}{Enhanced Type II\\ Port Selection} & $\left\{2,4\right\}$ &\tabincell{c}{phase, delay and \\amplitude (3bit)\\with compression} & $\begin{array}{l}
1 + \upsilon  + 2LM{}_\upsilon \upsilon  + 2{M_{\rm{nz}}},{N_3} \le 19;\\
2 + \upsilon  + 2LM{}_\upsilon \upsilon  + 2{M_{\rm{nz}}},{N_3} > 19
\end{array}$ & ${\cal O}\left( {2\upsilon L{ M_\upsilon }{d}} \right)$
\tabularnewline
\hline
\tabincell{c}{Further Enhanced \\Type II Port Selection} & $\left\{1,2,3,4,6\right\}$ & \tabincell{c}{phase, delay and \\amplitude (3bit)\\with enhanced \\compression} & $\begin{array}{l}
1 + 4LM\upsilon ,N = 2,\upsilon  \le 2;\\
1 + 2ML\upsilon  + 2{M_{\rm{nz}}},N = 2,\upsilon  > 2;\\
2 + 4LM\upsilon ,N = 4,\upsilon  \le 2;\\
2 + 2ML\upsilon  + 2{M_{\rm{nz}}},N = 4,\upsilon  > 2
\end{array}$ &  ${\cal O}\left( {2\upsilon {L^2}{M}} \right)$
\tabularnewline
\hline
\end{tabular}
\end{table*}
In our previous discussion, we elaborated on the codebooks and the corresponding PMI report mechanisms of all codebooks supported in 5G NR so far. The key properties of these codebooks are summarized in Table~\ref{tab codebook performance}, which is a comparison in terms of the number of reported spatial beams, the subband coefficient quantization manner, the feedback overhead, and the computational complexity. Note that the feedback overhead is quantified by the number of all reported coefficients and indicators through all subbands. The complexity refers to the precoding matrix calculating complexity for the gNB.

In fact, different codebooks might be adopted according to different system requirements and application scenarios. {Overall, the trade-off between performance and feedback overhead is the key point for codebook choice.} 

For example, in SU-MIMO, Type I Codebook may be sufficient due to its simplicity. {And it performs well in simple wireless propagation environments, such as LOS scenarios.} However, Type II Codebook and the succeeding codebooks support multiple beams and thereby may outperform Type I Codebook in MU-MIMO due to the better mitigation of multi-user interference. Moreover, in wideband massive MIMO, Enhanced Type II Codebook can effectively reduce the feedback overhead compared to Type II Codebook. In case of low feedback and complexity constraints at the UE side, port selection codebooks are more preferable than non-port selection codebooks. {And port selection codebooks offer better flexibility in terms of the precoder form constraint and the BS antenna topology.} Particularly, in wideband FDD massive MIMO, Further Enhanced Type II Codebook may be a better choice due to the reduced feedback overhead and increased system performance brought by the exploitation of partial channel reciprocity.

Nowadays, new multiple antenna technologies are emerging, and the application scenarios are extending. They call for suitable codebooks to accommodate specific scenarios and system requirements. In the following of the paper, we will discuss some unresolved challenges of codebooks for the future and some promising solutions to these challenges.

\subsection{Enhanced Codebook for mobile scenarios}
One of the major challenges in massive MIMO is the mobility problem. The shorter coherence time in mobility scenarios leads to a serious degradation on the spectral efficiency \cite{2019mobilityReport}. Recently, some research work focused on this dilemma from the theoretical perspective \cite{2020yinMobility,2021Karlman,2021dl_mobility,YongLiao2023}. In industry, the topic of mobility enhancement had been considered and discussed in R17 from the perspective of mobility management with a type of non-zero power (NZP) CSI-RS for mobility management, as well as synchronization signal block (SSB)-based handover between cells. In future R18 version or 5G-advanced, the enhancement of mobility performance  will be an integral part and has been added to the agenda \cite{2022mobilityMeeting}. However in the 5G NR standards, no codebook has ever been particularly designed for the non-negligible UE mobility, which causes serious deterioration of the system spectral efficiency. The main reason lies in the fast variation of the channel, and in particular, the Doppler of the paths. Unfortunately, the codebooks supported in R17 cannot solve this challenge. They are not designed to characterize the Doppler frequency shift of the multipath of the channel, nor can they report timely CSI for high mobility scenarios.

In order to design an enhanced codebook for mobility scenarios, we believe that three constraints should be considered. First, the codebook should characterize the Doppler frequency shift information of the channel. Second, the time-varying channel demands a timely CSI feedback framework. Introducing a channel prediction scheme in CSI report may be a solution. 
Third, the compatibility with the existing codebooks is also vital and will facilitate the implementation. The mobility enhanced codebook proposed in \cite{qin2022mobility} is a candidate, which provides an effective approach to obtain the CSI in high mobility scenarios by applying a joint-angle-delay-Doppler (JADD) channel prediction scheme. The core idea is to track the multipath Doppler frequency shifts with a few channel samples. Moreover, the timely CSI feedback is enabled by a partial reciprocity based wideband precoding scheme for the pilots and the feedback-based CSI prediction at the gNB.  

\subsection{Codebook for cell-free massive MIMO}
Recently, a distributed multiple antenna system or cell-free massive MIMO system has drawn much attention by academia \cite{chen2020wireless} and industry. Compared to cell-centric massive MIMO, cell-free massive MIMO aims to serve the UEs simultaneously through widely distributed access points (APs) instead of the centralized antenna array at the base station \cite{2017cellfree}. This cell-free massive MIMO mainly shows the advantage in exploiting diversity against shadow fading at the expense of high backhaul requirements. It may lead to performance improvements in respect of the coverage probability, the energy efficiency and the spectral efficiency \cite{nayebi2017precoding}.

In cell-free massive MIMO, the channel environments between the distributed APs and a certain UE are quite different. There is little correlation between the distributed BS antennas, which makes the CSI compression more challenging. In fact, the codebooks mentioned in this paper all rely on the channel structure, e.g., the spatial-domain structure of the multipath angular response and the frequency-domain structure of the multipath delay response. The structure makes the channel correlated and therefore compressible. In cell-free massive MIMO however, such structures are not available. Hence, it is more challenging to characterize the channel parameters of each path, and the current codebook framework may not be suitable for cell-free massive MIMO. 
Therefore, the major challenge of designing the codebook for cell-free massive MIMO lies in how to reduce the feedback overhead, which scales with the number of distributed antennas. 

\subsection{Codebook for Ultramassive MIMO System}
Nowadays, 6G is widely discussed and many emerging technologies are considered candidates to be used in 6G \cite{2021Pcd6G,zhang20196g}, 
including ultramassive MIMO, which helps to meet the extremely high rate acquirement. One of the most tricky problem of ultramassive MIMO is the CSI acquisition due to the massive antenna arrays. We believe that the challenges are mainly reflected in the following three aspects. First, the channel propagation nature is coherently changed. Due to the increasing antenna dimension, the channel radiating environment tends to exhibit a near-field effect. Hence, current codebook which is based on far-field radiating condition may fail to characterize the channel environment. Second, the dimension of the CSI and the precoder increase significantly. Therefore, the complexity of precoding and signal processing in ultramassive MIMO increase exponentially. Third, future codebooks for ultramassive MIMO should be easy to implement in real communication system. {Some state-of-the-art methods specialize in dealing with the overwhelming CSI, such as artificial intelligence (AI) \cite{2020AIWCM}, compressed sensing method \cite{2010ProcCS}, two-stage beamforming \cite{2022dualStage} and hybrid beamforming \cite{2020hybrid}. They are promising solutions to utramassive MIMO codebook design.} Nevertheless, how these methods would be standardized and deployed in real communication systems is a problem that needs to be solved in the future. 

\section{Conclusion}\label{sec7}
In this paper, we discussed the codebook evolution from the 3GPP standard point of view. We first summarized the timeline and trend of codebook evolution. The physical meanings of the codebook parameters were given for a better grasp on the codebooks. Then we elaborated on feedback scheme and the PMI format of all codebooks in 5G NR. We also compared the performance of the codebooks in respect of the number of supported beams, subband quantization and feedback manner, feedback overhead, and the complexity at the gNB. {The numerical results of the performance vs. feedback overhead were given for different codebooks in 5G NR.} Finally, the remaining issues of codebook design for high mobility scenarios were discussed, and the open problems of codebook for cell-free massive MIMO and ultramassive MIMO were raised.









\ifCLASSOPTIONcaptionsoff
  \newpage
\fi
\bibliographystyle{IEEEtran}
\bibliography{IEEEabrv,reference}
\begin{IEEEbiography}[{\includegraphics[width=1in,height=1.25in,clip,keepaspectratio]{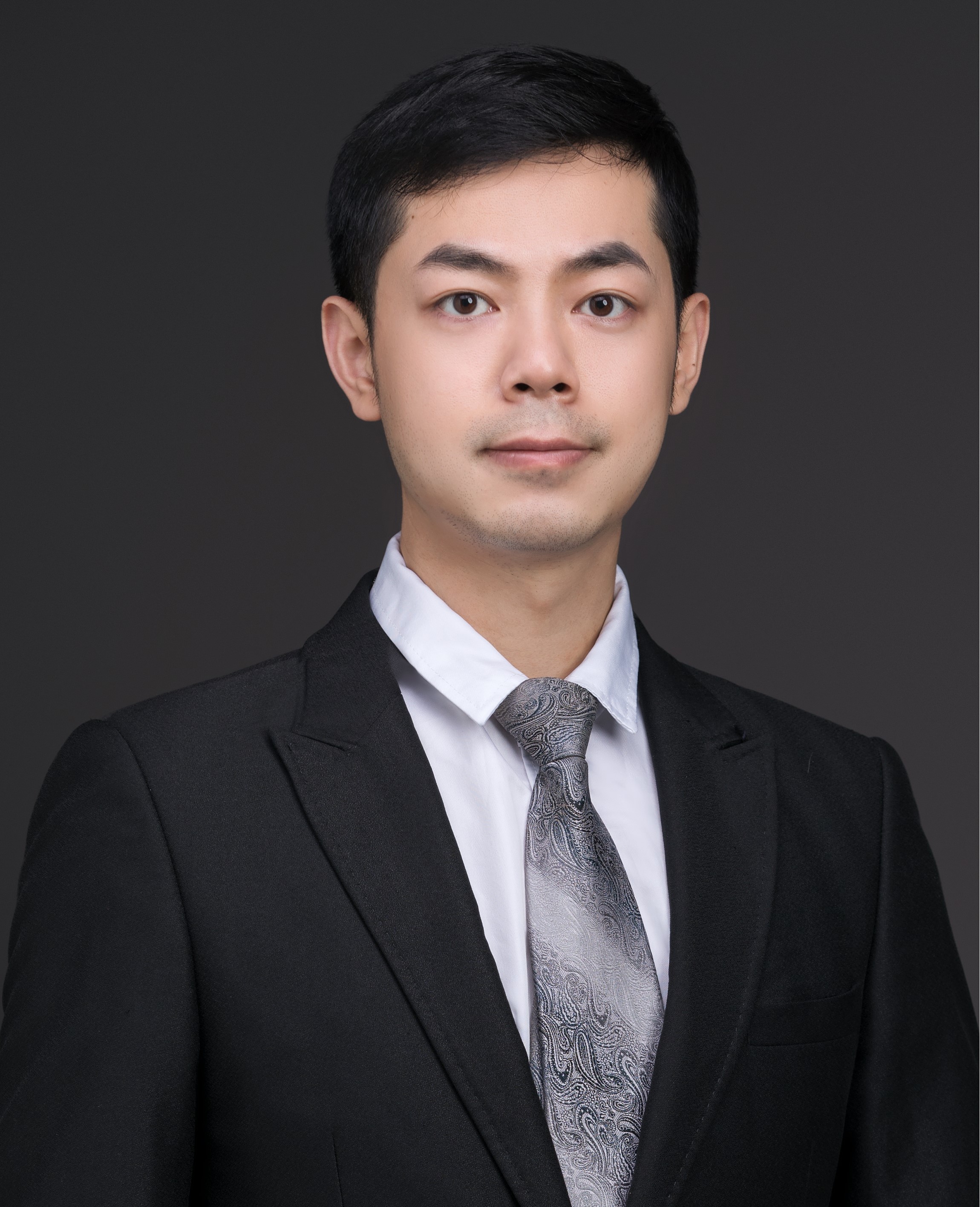}}]
{Ziao Qin} received the B.Sc. degree in Information Engineering from Beijing Institute of Technology, Beijing, China, in 2014. From 2014 to 2017, he works in industry in Beijing. Since 2018, he has been a graduate student at Huazhong University of Science and Technology, Wuhan, China. He is currently pursuing the Ph.D. degree in Information and Communications Engineering. His research interests include channel estimation, signal processing, codebook design, and beamforming for massive MIMO systems.
\end{IEEEbiography}

\begin{IEEEbiography}[{\includegraphics[width=1in,height=1.25in,clip,keepaspectratio]{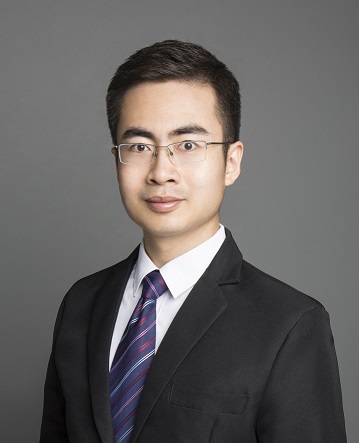}}]
{Haifan Yin} received the Ph.D. degree from T\'el\'ecom ParisTech in 2015. He received the B.Sc. degree in Electrical and Electronic Engineering and the M.Sc. degree in Electronics and Information Engineering from Huazhong University of Science and Technology, Wuhan, China, in 2009 and 2012 respectively. From 2009 to 2011, he has been with Wuhan National Laboratory for Optoelectronics, China, working on the implementation of TD-LTE systems as an R\&D engineer.
From 2016 to 2017, he has been a DSP engineer in Sequans Communications - an IoT chipmaker based in Paris, France. From 2017 to 2019, he has been a senior research engineer working on 5G standardization in Shanghai Huawei Technologies Co., Ltd., where he made substantial contributions to 5G standards, particularly the 5G codebooks. Since May 2019, he has joined the School of Electronic Information and Communications at Huazhong University of Science and Technology as a full professor. 
His current research interests include 5G and 6G networks, signal processing, machine learning, and massive MIMO systems. H. Yin was the national champion of 2021 High Potential Innovation Prize awarded by Chinese Academy of Engineering, a winner of 2020 Academic Advances of HUST, and a recipient of the 2015 Chinese Government Award for Outstanding Self-financed Students Abroad.
\end{IEEEbiography}

\end{document}